\documentclass[aip,reprint,longbibliography]{revtex4-1}
\usepackage{graphicx,color}
\usepackage{amsmath, amssymb}
\usepackage{longtable}
\usepackage{natbib}
\usepackage{bm}
\usepackage{multirow}
\allowdisplaybreaks[1]

\begin{document}
\title{First principles study of the vibronic coupling in positively charged C$_{60}^+$}
\author{Zhishuo Huang}
\email{zhishuohuang@gmail.com}
\affiliation{Theory of Nanomaterials Group, KU Leuven, Celestijnenlaan 200F, B-3001 Leuven, Belgium}
\author{Dan Liu}
\email{iamdliu@nwpu.edu.cn}
\affiliation{Institute of Flexible Electronics (IFE), Northwestern Polytechnical University, 127 West Youyi Road, Xi’an, 710072, Shaanxi, China}
\affiliation{Theory of Nanomaterials Group, KU Leuven, Celestijnenlaan 200F, B-3001 Leuven, Belgium}

\date{\today}

\begin{abstract}
Vibronic coupling parameters for C$_{60}^{+}$ were derived via DFT calculations with hybrid B3LYP and CAM-B3LYP functional,
based on which the static Jahn-Teller effect were analyzed.
The global minima of adiabatic potential energy surface (APES) shows a D$_{5d}$ Jahn-Teller deformation,
with stabilization energies of 110 and 129 meV (with B3LYP and CAM-B3LYP respectively),
which are two times larger than that in C$_{60}^-$,
suggesting the crucial role of the dynamical Jahn-Teller effect.
Present results enable us to assess the actual situation of dynamical Jahn-Teller effect in C$_{60}^{+}$ and excited C$_{60}$ in combination with the established parameters for C$_{60}^{-}$.
\end{abstract}

\maketitle

\section{Introduction}
Recent experimental confirmation of the presence of C$_{60}^+$ as interstellar materials \cite{Campbell2015, Campbell2016} renewed the interests in C$_{60}$ cations, 
and burst various spectroscopic and theoretical investigations on C$_{60}^+$ and related systems \cite{Spieler2017, Yamada2017, Strelnikov2018, Kaiser2018, Lykhin2019, ConfirmInterstellarC60plus2019}. 
It is known that C$_{60}$ in its charged and excited states exhibits complex Jahn-Teller (JT) effect due to its high symmetry ($I_h$) \cite{Chancey1997, Bersuker2006}. 
In particular, JT effect in C$_{60}^+$ is one of the most involved cases because of the five-fold degenerate highest occupied molecular orbitals (HOMOs) of C$_{60}$.
Toward understanding JT effect in C$_{60}^{n+}$ cations, spectroscopic \cite{Bruhwiler1997, Canton2002, Kern2013} and theoretical \cite{Ceulemans1990, Moate1996, DeLosRios1996, Manini2003, Lijnen2005, Hands2006a, Hands2006b, Hands2007, Ceulemans_S5} investigations have been piled up.

Realistic description of JT effect in positively charged C$_{60}$ relies on the combination of an adequate model and accurate enough vibronic coupling parameters.
Derivations of vibronic coupling parameters have been addressed \cite{Manini2001, Saito2002, Ramanantoanina2013}, and comprehensive sets of parameters have been estimated by density functional theory (DFT) calculations at local density approximation (LDA) level \cite{Manini2001, Ramanantoanina2013}. 
Nevertheless, in studies about C$_{60}^-$, it has been shown that LDA tends to underestimate the coupling parameters \cite{Iwahara2010}, while hybrid B3LYP functional is found to give closer parameters to the experimental data. 
Furthermore, a good agreement between B3LYP and GW approximation calculations \cite{Faber2011} supports the accuracy of hybrid functional in studies of C$_{60}$. 
Besides, one recent study showed that B3LYP with long-range interaction correction, CAM-B3LYP, could improve the accuracy of vibronic parameters in C$_{60}^-$ with respect to experimental data, indicating CAM-B3LYP could give vibronic parameters much closer to the real situation\cite{Huang_2019_C60t1g}.
Therefore, it is desired to derive the coupling parameters at a better level than LDA for accurate description of C$_{60}$ cations.

In this work, we derived orbital vibronic coupling parameters for C$_{60}^{+}$ via DFT calculations with both B3LYP and CAM-B3LYP hybrid functionals. 
These obtained vibronic coupling parameters were compared with the previous data obtained by LDA calculations.
Based on these parameters, the adiabatic potential energy surface (APES) was analyzed, and the symmetry of JT deformed C$_{60}^+$ as well as static JT energies were established.
\section{Vibronic Hamiltonian}
\label{Sec:H}
The highest occupied molecular orbitals (HOMOs) of C$_{60}$ with $I_h$ symmetry \cite{C60_report_first} are characterized by five-fold degenerate $h_u$ irreducible representation.
 According to selection rule, these orbitals linearly couple to mass-weighted normal vibrational modes involved in the symmetric product of $h_u$ representation \cite{Jahn1937}:
\begin{eqnarray}
 [h_u \otimes h_u] = a_g \oplus g_g \oplus 2 h_g.
\label{Eq:selection}
\end{eqnarray}
Among them, $g_g$ and $h_g$ modes are JT active, while $a_g$ is not because is does not change the symmetry of molecules.
Thus, taking the equilibrium structure of neutral C$_{60}$ as the reference, $H \otimes (a \oplus g \oplus 2h)$ JT Hamiltonian for C$_{60}^+$ is expressed as \cite{Ceulemans1990, Chancey1997, Bersuker2006}
\begin{eqnarray}
 H &=& H_{0} + H_{ \text{JT}}, \\
\label{Eq:H}
 H_0 &=& \frac{1}{2} \left(p_a^2 + \omega_a^2 q_{a}^2 \right) + V_a q_{a} + \sum_{\gamma = a, x, y, z} \frac{1}{2}\left(p_{g\gamma}^2 + \omega_g^2 q_{g \gamma}^2\right) \nonumber \\
 &+& \sum_{\gamma = \theta, \epsilon, \xi, \eta, \zeta} \frac{1}{2}\left(p_{h\gamma}^2 + \omega_h^2 q_{h \gamma}^2\right),
\label{Eq:U0}
\\
 H_\text{JT} &=&
  V_g \sum_{\gamma = a, x, y, z} \hat{C}_{g\gamma} q_{g\gamma}+ \frac{\sqrt{5}V_{1 h}}{2}\sum_{\gamma = \theta, \epsilon, \xi, \eta, \zeta} \hat{C}_{1h\gamma} q_{1h\gamma}\nonumber \\
&+&\frac{\sqrt{5} V_{2 h}}{2} \sum_{\gamma = \theta, \epsilon, \xi, \eta, \zeta} \hat{C}_{2h\gamma} q_{2h\gamma}
\label{Eq:HJT}
\end{eqnarray}

\noindent where $\omega_\Gamma$ ($\Gamma = a_g, g_g, h_g$) are vibration frequencies, $q_\Gamma$ are mass-weighted normal coordinates \cite{Inui1990}, 
$V_\Gamma$ is vibronic coupling parameter for $\Gamma$ mode, and $\hat{C}_{\Gamma\gamma}$ ($\gamma$ = a, x, y, z, $\theta$, $\epsilon$, $\xi$, $\eta$, $\zeta$) are the Clebsch-Gordan coefficients, which are taken from Ref. \cite{Liu2018b} and listed in Appendix A. 
A coefficient $\sqrt{5}/2$ is multiplied to vibronic couplings terms of $h_g$ modes so that JT energy becomes:
\begin{eqnarray}
\label{Eq:EJT_stabilization}
    E_{g}^\text{JT}=-\frac{V_{g}^2}{2\omega^{2}_{g}}, \quad
    E_{nh}^\text{JT}=-\frac{V_{nh}^2}{2\omega^{2}_{h}}, \quad (n=1,2).
\end{eqnarray}
There are two vibronic couplings to one $h_g$ mode because $h_g$ representation appears twice in selection rule, Eq. (\ref{Eq:selection}).
The basis of vibronic Hamiltonian matrices are $H_u$ electronic states of C$_{60}^+$ in the order of $|H_u \theta\rangle$, $|H_u \epsilon\rangle$, $|H_u \xi\rangle$, $|H_u \eta\rangle$, $|H_u \zeta\rangle$.
For $h_g$ and $h_u$ representations, $d$ orbital type basis are used, and hence, $\theta, \epsilon, \xi, \eta, \zeta$ transform as $(2z^2-x^2-y^2)/\sqrt{6}$, $(x^2-y^2)/\sqrt{2}$, $\sqrt{2}yz$, $\sqrt{2}zx$, $\sqrt{2}xy$, respectively, under rotation.
Although C$_{60}$ has two $a_g$, six and eight sets of $g_g$ and $h_g$ modes, respectively, and indices distinguishing them are not explicitly written in Eq. (\ref{Eq:H}) for simplicity.
Phase factors of mass-weighted normal modes are the same as those in the Supplemental Materials of Ref. \cite{Liu2018b}. 
Since the equilibrium geometry of C$_{60}$ is chosen as the reference structure of C$_{60}^+$, vibronic coupling parameters of totally symmetric modes are also nonzero.
JT energy by totally symmetric deformation is
\begin{eqnarray}
 E_a &=& -\frac{V_a^2}{2\omega_a^2}.
\label{Eq:Ea}
\end{eqnarray}

In many other literatures, like the work of A. Ceulemans\cite{Ceulemans1990}, linear combinations of real $d$-type functions, $d_{z^2}, d_{x^2-y^2}$, are used, which could be transformed into irreducible representation in this work by
\begin{eqnarray}
 q_\theta =\sqrt{\frac{5}{8}} Q_\theta + \sqrt{\frac{3}{8}} Q_\epsilon, \quad
 q_\epsilon=\sqrt{\frac{3}{8}} Q_\theta - \sqrt{\frac{5}{8}} Q_\epsilon,
\end{eqnarray}
where $Q_\theta$ and $Q_\epsilon$ are normal coordinates in Ref. \cite{Ceulemans1990}, while $q_\theta$ and $q_\epsilon$ are normal coordinates in this work, resulting in the relation between coupling parameters $V_{1h}$ and $V_{2h}$ in this work and $F_{Hb}$ and $F_{Ha}$ defined in Ref. \onlinecite{Ceulemans1990}:
\begin{eqnarray}
 V_{1h} = \frac{F_{Hb}}{4\sqrt{5}}, \quad V_{2h} = \frac{F_{Ha}}{12\sqrt{5}}.
\end{eqnarray}
The modification introduced here is to treat the JT Hamiltonian in a framework consistent to the standard one in C$_{60}^-$ \cite{Chancey1997}.

\section{Results}
\subsection{Orbital vibronic coupling parameters}
\label{Sec:V}
Vibronic coupling constants of C$_{60}$ have been most intensively investigated in the case of C$_{60}^-$, and coupling constants have been derived by various methods.
By definition, vibronic coupling parameters for JT active $h_g$ modes, $V_{h_g}^-$, can be derived by \cite{Iwahara2010}

\begin{eqnarray}
 V_{h_g}^- &=& -\left.\frac{\partial E_{z}^-(\textbf{q}_{h_g})}{\partial q_{h_g\theta}}\right| _{\textbf{q}_{h_g}=\textbf{0}},
\end{eqnarray}

where $E_z^-(\textbf{q}_{h_g}) = \langle T_{1u}z| \hat{H}^- |T_{1u}z\rangle$, $|T_{1u}z\rangle$ is $t_{1u}z$ electronic state of C$_{60}^-$, $\hat{H}^-$ is Hamiltonian for C$_{60}^-$, and $\textbf{q}_{h_g}=\textbf{0}$ indicates equilibrium structure of C$_{60}$.
Because of symmetry, contributions from occupied orbitals are zero, and only partially filled $t_{1u}$ orbital level contribute to vibronic couplings.
In the case of C$_{60}^-$, the nature of $t_{1u}$ orbitals do not differ from that of neutral C$_{60}$: 
although $t_{1u}$ orbitals are mixed with the other $t_{1u}$ orbitals, the mixing is very small due to high symmetry, and orbitals are separated from each other by large orbital energy gaps.
Consequently, gradients of total energy can be approximated by those of orbital energy levels with respect to normal modes of neutral C$_{60}$.
Indeed, in the case of C$_{60}^-$, these two approaches give very close results \cite{Janssen2010, Iwahara2010, Faber2011, Liu2018b}.

Similar situation is expected in C$_{60}^+$: the nature of the $h_u$ orbital does not change by adding one hole to the same molecular structure.
Thus, orbital vibronic coupling parameters of neutral C$_{60}$ can be used to express vibronic coupling parameters of C$_{60}^+$.
Since C$_{60}^+$ has one hole in $h_u$ HOMOs, it is convenient to perform particle-hole transformation \cite{Fetter}, 
under which, the sign of orbital vibronic coupling parameter for one electron in HOMOs of C$_{60}$ should be inverted for that in the case of one hole in HOMOs.
Therefore,
\begin{eqnarray}
    V_{\Gamma} = -v_{\Gamma},
\label{Eq:Vv}
\end{eqnarray}
where $v_{\Gamma}$ is one orbital vibronic coupling parameter for C$_{60}$ and $V_\Gamma$ is the parameter for C$_{60}^+$.

Orbital vibronic coupling parameters were calculated using frozen-phonon approach.
Orbital energy levels of distorted C$_{60}$ are fitted to the eigenvalues of JT Hamiltonian matrix (Eq. (\ref{Eq:HJT})).
In the present case, ${h_g(\mu)\epsilon}$ and ${g_g(\mu)a}$ deformations are used because diagonalizing the model Hamiltonian is easier.
For DFT calculations, a triple-zeta basis set [6-311G(d)] was employed for both B3LYP and CAM-B3LYP functionals within Gaussian \cite{g16}.
Some fittings are shown in Fig. \ref{Fig:HOMO} (see Supplemental Materials for other fittings).
Black points indicate DFT levels originating from HOMOs, with gray lines for energy level calculated from model Hamiltonian.
Derived orbital vibronic coupling parameters are shown in Table \ref{Table:V}. From this table, JT stabilization energies for different JT active modes are improved about 30 \% with CAM-B3LYP compared with these with B3LYP.
One should note that there is almost no nonlinear splitting due to vibronic effect in HOMO levels, indicating weak quadratic or higher vibronic couplings as in the case of C$_{60}^-$ \cite{Liu2018b}.
This guarantees the validity of linear vibronic model (Eq. (\ref{Eq:H})) for the description of JT effect of C$_{60}$ cations.

\begin{figure}
\begin{center}
\includegraphics[height=5cm]{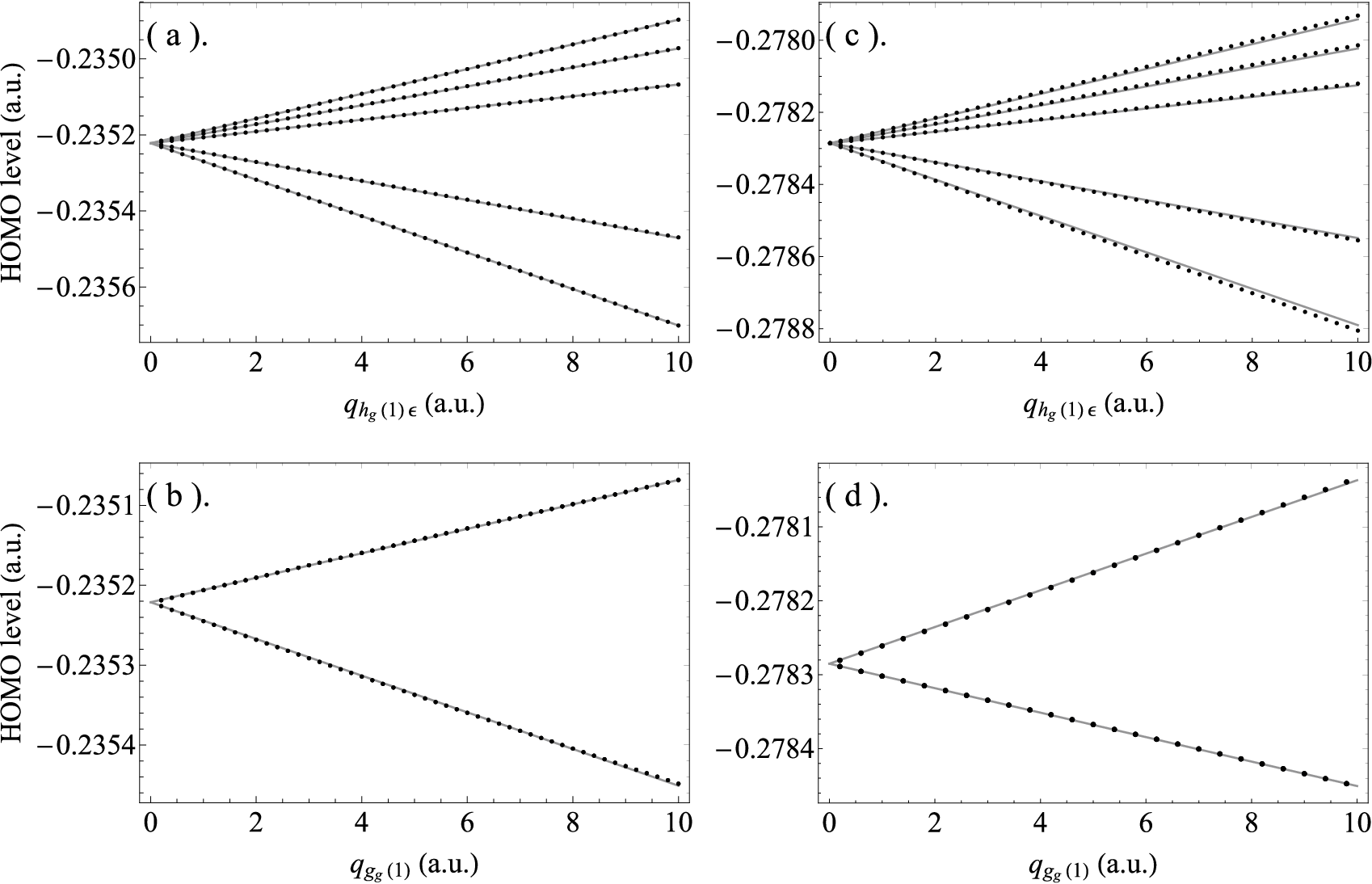}
\caption{
The JT splitting of the HOMO levels with respect to $q_{h_g(1)\epsilon}$ ((a) and (c)) and $q_{g_g(1)a}$ ((b) and (d)) deformations (in atomic unit). 
(a) and (b) are calculated with B3LYP functional, while (c) and (d) are with CAM-B3LYP functional.
The black points and gray lines indicate the DFT values and model energy, respectively.}
\label{Fig:HOMO}
\end{center}
\end{figure}

\begin{table*}[!htb]
\begin{center}
\caption{The frequencies $\omega_\Gamma$ (cm$^{-1}$), orbital vibronic coupling parameters $v_\Gamma$ (10$^{-4}$ a.u.), and stabilization energies $E_\Gamma$ (meV) for the $a_g$, $g_g$, and $h_g$ modes. 
The data calculated with B3LYP for LUMOs are taken from Ref. \cite{Liu2018b}.
$g_\Gamma = V_\Gamma/\sqrt{\hslash \omega_\Gamma^3}$ is dimensionless vibronic coupling parameter.
}

  \label{Table:V}
  \resizebox{\textwidth}{!}{\begin{tabular}{ccccccccccccccc}
  \hline
  & \multicolumn{7}{c}{B3LYP} & \multicolumn{7}{c}{CAM-B3LYP} \\
$\mu$ & $\omega_{\Gamma}$ & \multicolumn{2}{c}{$v_\Gamma$} & \multicolumn{2}{c}{$g_\Gamma$} & \multicolumn{2}{c}{$E_\Gamma$} &  $\omega_{\Gamma}$ & \multicolumn{2}{c}{$v_\Gamma$} & \multicolumn{2}{c}{$g_\Gamma$} & \multicolumn{2}{c}{$E_\Gamma$}\\
  \hline
      &                   & 1 & 2 & 1 & 2 & 1 & 2  & &1 & 2 & 1 & 2 & 1 & 2\\
\hline
\multicolumn{15}{c}{HOMO} \\
\hline
$a_{g1}$& 497&\multicolumn{2}{c}{$-0.121$}&\multicolumn{2}{c}{$-0.112$}&\multicolumn{2}{c}{0.389}& 506&\multicolumn{2}{c}{$-0.118$}&\multicolumn{2}{c}{$-0.106$}&\multicolumn{2}{c}{0.355}\\
$a_{g2}$&1498&\multicolumn{2}{c}{1.040}   &\multicolumn{2}{c}{0.184}   &\multicolumn{2}{c}{3.159}&1527&\multicolumn{2}{c}{1.369}   &\multicolumn{2}{c}{0.236}   &\multicolumn{2}{c}{5.268}\\
 \hline
$g_{g1}$& 481&\multicolumn{2}{c}{$-0.375$}&\multicolumn{2}{c}{$-0.366$}&\multicolumn{2}{c}{3.984}& 502&\multicolumn{2}{c}{$-0.405$}&\multicolumn{2}{c}{$-0.370$}&\multicolumn{2}{c}{4.265}\\
$g_{g2}$& 584&\multicolumn{2}{c}{$-0.101$}&\multicolumn{2}{c}{$-0.074$}&\multicolumn{2}{c}{0.196}& 586&\multicolumn{2}{c}{$-0.207$}&\multicolumn{2}{c}{$-0.150$}&\multicolumn{2}{c}{0.815}\\
$g_{g3}$& 768&\multicolumn{2}{c}{0.923}   &\multicolumn{2}{c}{0.446}   &\multicolumn{2}{c}{9.466}& 787&\multicolumn{2}{c}{1.083}   &\multicolumn{2}{c}{0.504}   &\multicolumn{2}{c}{12.400}\\
$g_{g4}$&1092&\multicolumn{2}{c}{$-1.281$}&\multicolumn{2}{c}{$-0.365$}&\multicolumn{2}{c}{9.019}&1092&\multicolumn{2}{c}{$-1.583$}&\multicolumn{2}{c}{$-0.451$}&\multicolumn{2}{c}{13.781}\\
$g_{g5}$&1335&\multicolumn{2}{c}{0.541}   &\multicolumn{2}{c}{0.114}   &\multicolumn{2}{c}{1.076}&1356&\multicolumn{2}{c}{0.739}   &\multicolumn{2}{c}{0.152}   &\multicolumn{2}{c}{1.947}\\
$g_{g6}$&1540&\multicolumn{2}{c}{1.477}   &\multicolumn{2}{c}{0.251}   &\multicolumn{2}{c}{6.029}&1576&\multicolumn{2}{c}{1.571}   &\multicolumn{2}{c}{0.258}   &\multicolumn{2}{c}{6.509}\\
  \hline
$h_{g1}$& 266&  0.690  & $-0.080$ &   1.635  & $-0.190$ & 44.099 &  0.593 &  271&  0.727  & $-0.090$ &   1.674  & $-0.206$ & 47.120 & 0.714 \\
$h_{g2}$& 439&$-0.508$ & $-0.290$ & $-0.568$ & $-0.324$ &  8.776 &  2.860 &  450&$-0.558$ & $-0.363$ & $-0.601$ & $-0.391$ & 10.063 & 4.269 \\
$h_{g3}$& 726&$-0.043$ &   0.977  & $-0.023$ &   0.514  &  0.023 & 11.869 &  745&$-0.061$ &   1.053  & $-0.031$ &   0.532  &  0.043 &13.086 \\
$h_{g4}$& 786&  0.923  & $-0.059$ &   0.431  & $-0.027$ &  9.038 &  0.037 &  802&  0.930  & $-0.139$ &   0.421  & $-0.063$ &  8.821 & 0.198 \\
$h_{g5}$&1125&$-0.101$ & $-0.495$ & $-0.028$ & $-0.135$ &  0.053 &  1.269 & 1148&$-0.074$ & $-0.565$ & $-0.020$ & $-0.150$ &  0.028 & 1.591 \\
$h_{g6}$&1269&  0.965  &   0.272  &   0.220  &   0.062  &  3.790 &  0.301 & 1300&  0.986  &   0.025  &   0.216  &   0.005  &  3.768 & 0.002 \\
$h_{g7}$&1443&  2.860  &   1.487  &   0.537  &   0.279  & 25.745 &  6.960 & 1480&  3.425  &   2.010  &   0.618  &   0.363  & 35.072 &12.081 \\
$h_{g8}$&1607&  2.721  & $-1.542$ &   0.434  & $-0.246$ & 18.790 &  6.034 & 1663&  3.217  & $-1.701$ &   0.488  & $-0.258$ & 24.523 & 6.854 \\
  \hline
\multicolumn{15}{c}{LUMO} \\
  \hline
$a_{g1}$& 497&\multicolumn{2}{c}{$-0.264$}&\multicolumn{2}{c}{$-0.245$}&\multicolumn{2}{c}{ 1.849}& 506&\multicolumn{2}{c}{$-0.253$}&\multicolumn{2}{c}{$-0.228$}&\multicolumn{2}{c}{ 1.629}\\
$a_{g2}$&1498&\multicolumn{2}{c}{$-2.380$}&\multicolumn{2}{c}{$-0.422$}&\multicolumn{2}{c}{16.543}&1527&\multicolumn{2}{c}{$-2.921$}&\multicolumn{2}{c}{$-0.503$}&\multicolumn{2}{c}{23.971}\\
 \hline
$h_{g1}$& 266& \multicolumn{2}{c}{0.192}& \multicolumn{2}{c}{0.455} & \multicolumn{2}{c}{ 3.415} & 271& \multicolumn{2}{c}{0.209} &\multicolumn{2}{c}{0.481} &\multicolumn{2}{c}{ 3.884} \\
$h_{g2}$& 439& \multicolumn{2}{c}{0.450}& \multicolumn{2}{c}{0.503} & \multicolumn{2}{c}{ 6.886} & 450& \multicolumn{2}{c}{0.456} &\multicolumn{2}{c}{0.491} &\multicolumn{2}{c}{ 6.735} \\
$h_{g3}$& 726& \multicolumn{2}{c}{0.754}& \multicolumn{2}{c}{0.396} & \multicolumn{2}{c}{ 7.069} & 745& \multicolumn{2}{c}{0.849} &\multicolumn{2}{c}{0.429} &\multicolumn{2}{c}{ 8.512} \\
$h_{g4}$& 786& \multicolumn{2}{c}{0.554}& \multicolumn{2}{c}{0.259} & \multicolumn{2}{c}{ 3.256} & 802& \multicolumn{2}{c}{0.575} &\multicolumn{2}{c}{0.260} &\multicolumn{2}{c}{ 3.367} \\
$h_{g5}$&1125& \multicolumn{2}{c}{0.766}& \multicolumn{2}{c}{0.209} & \multicolumn{2}{c}{ 3.038} &1148& \multicolumn{2}{c}{0.827} &\multicolumn{2}{c}{0.219} &\multicolumn{2}{c}{ 3.402} \\
$h_{g6}$&1269& \multicolumn{2}{c}{0.578}& \multicolumn{2}{c}{0.132} & \multicolumn{2}{c}{ 1.360} &1300& \multicolumn{2}{c}{0.513} &\multicolumn{2}{c}{0.113} &\multicolumn{2}{c}{ 1.019} \\
$h_{g7}$&1443& \multicolumn{2}{c}{2.099}& \multicolumn{2}{c}{0.394} & \multicolumn{2}{c}{13.867} &1480& \multicolumn{2}{c}{2.553} &\multicolumn{2}{c}{0.461} &\multicolumn{2}{c}{19.492} \\
$h_{g8}$&1607& \multicolumn{2}{c}{2.043}& \multicolumn{2}{c}{0.326} & \multicolumn{2}{c}{10.592} &1663& \multicolumn{2}{c}{2.325} &\multicolumn{2}{c}{0.352} &\multicolumn{2}{c}{12.808} \\
\hline
  \end{tabular}}
\end{center}
\end{table*}

\subsection{Static Jahn-Teller effect}
Vibronic coupling lifts degeneracy with the deformation keeping one of the highest subgroup symmetries \cite{Liehr1963I}, resulting in six $D_{5d}$ and ten $D_{3d}$ minima \cite{Ceulemans1990}, as there are six $C_5$ and ten $C_3$ axes in C$_{60}$.
Thus, based on present basis, using symmetry adapted deformations \cite{Hands2006b}, deformations for $D_{5d}$ and $D_{3d}$ minima are expressed by
\begin{eqnarray}
\label{Eq:Deformation_D5d}
    {\textbf{q}}_{h_g}^{D_{5d}} &=& q_{h_{g}}\left(\frac{\phi^{2}}{2\sqrt{5}},\frac{\phi^{-1}}{2}\sqrt{\frac{3}{5}},0,\sqrt{\frac{3}{5}},0\right),
\nonumber\\
    {\textbf{q}}_{g_g}^{D_{5d}} &=& q_{g_{g}}\left(0,0,0,0\right),
\end{eqnarray}
and
\begin{eqnarray}
\label{Eq:Deformation_D3d}
    {\textbf{q}}_{h_g}^{D_{3d}} &=& q_{h_{g}}\left(-\frac{\phi^{-1}}{2},\frac{\phi^{2}}{2\sqrt{3}},0,\frac{1}{\sqrt{3}},0\right), \nonumber\\
    {\textbf{q}}_{g_g}^{D_{3d}} &=& q_{g_{g}}\left(\frac{1}{\sqrt{6}},0,-\sqrt{\frac{5}{6}},0\right),
\end{eqnarray}
respectively\cite{Liu2018b}.
Substituting these symmetrized deformations, Eqs. (\ref{Eq:Deformation_D5d}) and (\ref{Eq:Deformation_D3d}), into potential terms of model Hamiltonian (kinetic energy term is ignored), and then diagonalizing model Hamiltonian, we obtain the lowest adiabatic potential energies as
\begin{eqnarray}
    U^{D_{5d}} &=& \frac{1}{2}\omega^{2}_{h}q^{2}_{h}+V_{1h}q_{h},
\nonumber\\
    U^{D_{3d}} &=& \frac{1}{2}\omega^{2}_{g}q^{2}_{g}+\frac{1}{2}\omega^{2}_{h}q^{2}_{h}+\frac{1}{3}\left(2V_{g}q_{g}+\sqrt{5}V_{2h}q_{h}\right),
\label{Eq:EJT}
\end{eqnarray}
for $D_{5d}$ and $D_{3d}$ deformations, respectively.
Furthermore, these global minima energies could be expressed in terms of stabilization energies (Eq. (\ref{Eq:EJT_stabilization})), as\cite{Ceulemans1990}
\begin{eqnarray}
\label{Eq:EJT_D3d_D5d}
   E_\text{JT}^{D_{5d}} &=& -E_{1h}^\text{JT}, \nonumber\\
   E_\text{JT}^{D_{3d}} &=& -\frac{1}{9}\left(4E_{g}^\text{JT}+5E_{2h}^\text{JT}\right).
\end{eqnarray}
JT stabilization energies of C$_{60}^+$ are obtained using these equations and the calculated vibronic coupling parameters.
When treating C$_{60}^+$, we have to sum up contributions from all $g_g$ and $h_g$ modes.
Manini {\it et al}. also have derived vibronic coupling parameters and JT stabilization energies in the same way with LDA. \cite{Manini2001}
Besides, there is always the stabilization due to the totally symmetric modes (Eq. (\ref{Eq:Ea})).

JT stabilization energies of C$_{60}^+$ have been calculated with different methods with various functionals.
These methodologies to derive JT energies are classified into four.
We denote the present method (I).
In the second method, JT stabilization energy is directly obtained from the energy difference between high- and low-symmetric structures.
This method was employed by Lykhin {\it et al}. \cite{Lykhin2019} with B3LYP.
The third and the fourth methods, (III) and (IV), are called interaction mode approach \cite{Khlopin1978, Bersuker1989} and intrinsic distortion path approach \cite{MD_DFT_Bruyndonckx_1997, MD_DFT_Zlatar_2009, IDP_Zlatar_2010, Zlatar_2019}, respectively.
In both methods, vibronic coupling parameters or JT energies are extracted from optimized geometry with each subgroup of $I_h$.
The interaction mode induces deformation along JT minima from high-symmetric coordinates and is expressed by a linear combination of normal modes of high-symmetric C$_{60}$.
The coefficients of the linear combinations contain the information of the vibronic coupling.
By expanding the difference of JT deformed and high-symmetry geometries, $\Delta \textbf{R}_\text{JT}$, with the eigen modes of mass-weighted normal modes $\textbf{e}_i$ \cite{Inui1990} at high-symmetry structure as
\begin{eqnarray}
 \frac{V_i}{\omega_i^2} = c_i \sqrt{M} \Delta \textbf{R}_\text{JT}.\textbf{e}_{i},
\end{eqnarray}
the vibronic coupling parameters $V_i$ could be obained.
The frequencies $\omega_i$ are obtained from first principles calculations and $M$ is mass of carbon, and the coefficients $c_i$ depend on the structure of the JT interaction matrix.
This approach was used in Refs. \cite{Ramanantoanina2013, Muya2013}.
On the other hand, within the intrinsic distortion path analysis, the high-symmetric structure is expressed by the linear combination of the eigen vectors from low-symmetric structure.
Combining the vibronic coupling parameters derived from the deformation, $V'_i$, and the frequencies at the low-symmetric structure, $\omega'_i$, the JT energy could be written as $E_\text{JT} = \sum_i V_i^{\prime 2}/(2\omega_i^{\prime 2})$.
The last method was applied to C$_{60}^+$ in Ref. \cite{Ramanantoanina2013}.

JT stabilization energies in this work, as well as those from previous studies, are shown in Table \ref{Table:EJT}, 
from which we could see that CAM-B3LYP could enhance JT stabilization energies for $D_{5d}$ and $D_{3d}$ minima by 17\% and 30\% respectively compared to that with B3LYP.
And JT stabilization energies obtained with both B3LYP and CAM-B3LYP are larger than those from LDA or PBE-related functionals.
In particular, data calculated with LDA from Manini {\it et al}.,\cite{Manini2001} is only about 60 \% of the present data.
The underestimation of JT energy by LDA method is consistent with the situation in C$_{60}^-$ \cite{Iwahara2010}.
However, since with LDA methods (II) and (III) give similar data as the one by method (I), \cite{Manini2001} the difference of these methods would not be the origin of discrepancies seen in B3LYP data.
Thus, a possible reason is that the deformed geometry with B3LYP functional in Ref. \cite{Ramanantoanina2013} is the one at a local minima.
For B3LYP, Ref. \cite{Lykhin2019} and present work give much close results, whereas the former is larger than the latter by 11 meV.
So such a difference is expected to come from that JT stabilization energies contributed by totally symmetric modes, Eq. (\ref{Eq:Ea}), are included in Ref. \cite{Lykhin2019}.
However, JT stabilization energy contributed from $a_g$ modes in this work is about 3.5 meV (Table \ref{Table:V}), which is still smaller than 11 meV.
The underestimation of vibronic coupling of totally symmetric modes may come from the lack of contributions from occupied orbitals and change of frequencies.
In general, only the partially filled frontier orbitals contribute to the vibronic coupling to the JT modes due to the symmetry \cite{SatoJT}, whereas all occupied orbitals do contribute to vibronic coupling to the totally symmetric modes \cite{SatoA}.
Indeed, in the case of a planar molecule, picene, orbital vibronic coupling parameters for totally symmetric modes differ by 10-20 \% from those obtained by fitting the gradients of total system energy \cite{Sato2012}.
Frequencies can be changed due to the removal of electron by 5-15 \% \cite{Matsuda2018}.
Another possible origin of this discrepancy is nonlinear vibronic coupling, however, such coupling is much weaker than linear vibronic coupling as in the case of C$_{60}$ anions \cite{Liu2018b}.

\begin{table}[h]
\begin{center}
\caption{
    Jahn-Teller energies of C$_{60}^+$ (meV) for $D_{5d}$ and $D_{3d}$ minima of the APES, respectively.
}
\label{Table:EJT}
\begin{tabular}{ccccc}
    \hline
    Functional & Method  & $D_{5d}$ & $D_{3d}$ & Ref\\
    \hline
    B3LYP  & (I)     & 110& 30  &Present         \\
 CAM-B3LYP & (I)     & 129& 39  &Present         \\
    LDA    & (I)     & 69 & 22  &\cite{Manini2001}       \\
    B3LYP  & (II)    & 121& -   &\cite{Lykhin2019}\\
    LDA    & (III)   & 74 & 27  &\cite{Ramanantoanina2013} \\
    OPBE   & (III)   & 74 & 28  &\cite{Ramanantoanina2013}\\
    B3LYP  & (III)   & 80 & 32  &\cite{Ramanantoanina2013}\\
    PBE    & (III)   & 74 & 28  &\cite{Muya2013}\\
    LDA    & (IV)    & 72 & 20  &\cite{Ramanantoanina2013}\\
    OPBE   & (IV)    & 74 & 21  &\cite{Ramanantoanina2013}\\
    B3LYP  & (IV)    & 94 & 25  &\cite{Ramanantoanina2013}\\
    \hline
\end{tabular}
\end{center}
\end{table}

Besides these works, we note that Kern {\it et al} \cite{Kern2013} have optimized the structure to simulate infrared (IR) absorption spectrum with BP86 functional, but JT energy was not derived.

\section{Discussion}
To fully reveal the molecular nature of C$_{60}^+$, non-adiabatic dynamical Jahn-Teller effect is crucial.
The most straightforward way would be exact diagonalizing the molecular Hamiltonian, which fully quantizes both nuclear and electronic coordinates, nevertheless, it is not practical.
To partly overcome this difficulty, combining Jahn-Teller model Hamiltonian (\ref{Eq:H}) with accurate enough vibronic coupling parameters is indispensable to derive low-energy states.

Vibronic coupling parameters of C$_{60}^+$ have been derived by using DFT calculations with LDA \cite{Manini2001, Ramanantoanina2013} and B3LYP \cite{Saito2002} functionals.
As discussed in Sec. \ref{Sec:H}, model Hamiltonian is described by one vibronic coupling parameter for each $a_g$ and $g_g$ mode, and two parameters for each $h_g$ mode.
Both parameters for the $h_g$ modes, $V_{1h}$ and $V_{2h}$, have been derived only in Ref. \cite{Manini2001} and Ref. \cite{Ramanantoanina2013}, while they have not in Ref. \cite{Saito2002}.
Ramanantoanina {\it et al}. \cite{Ramanantoanina2013} shows that the magnitudes of derived coupling parameters obtained from the gradient of HOMO levels \cite{Manini2001} and those from adiabatic potential energy surface agree well with each other, which has been also seen in C$_{60}^-$ \cite{Iwahara2010, Janssen2010, Liu2018b}.
Thus, the present orbital vibronic coupling parameters derived from C$_{60}$ must be close to the parameters for C$_{60}^+$ derived based on the definition.

The accuracy of LDA data have been discussed \cite{Ramanantoanina2013, Ponzellini} based on the comparison between experimental photoelectron spectra (PES) \cite{Canton2002} and those from numerical simulation \cite{Manini2003}.
Indeed, PES is very useful to establish vibronic coupling parameters, whereas the simulation of PES requires high accuracy both in theoretical simulation and experiments.
In the case of C$_{60}^-$, vibronic coupling parameters derived from broad PES at high-temperature \cite{Gunnarsson1995} have been proved to be overestimated by the simulation of high-resolution PES spectra \cite{Iwahara2010}.
Furthermore, it was also found that the error bar of vibronic coupling parameters derived from broad PES is very large \cite{Iwahara2010}.
Thus, the derivation of accurate coupling parameters is only possible if we have high-resolution PES spectra measured at low-temperature \cite{Wang2005, PES_C60-}.
In the case of C$_{60}^+$, as pointed out by Manini {\it et al}. \cite{Manini2003}, experimental PES is broad and fine structure of low-energy region due to vibronic coupling is completely smeared out, which prevents the direct comparison between theory and experiment.
Moreover, in the case of PES of C$_{60}^+$, theoretical ratio of the second strongest peak to the strongest one is smaller than those of experimental data, implying the underestimation of vibronic coupling by LDA.
From this point, the present data larger than the LDA data by 40 \% would give better agreement.

The quality of B3LYP calculations has been checked in C$_{60}$ anions by comparing theoretical and experimental data in previous studies.
Besides the good agreement between coupling parameters from B3LYP calculations and high-resolution PES \cite{Iwahara2010}, the good quality of B3LYP calculations also has been confirmed from N\'{e}el temperature \cite{Iwahara2013}, spin gap \cite{Liu2018a}, and the explanation for the origin of temperature evolution of infrared spectra \cite{Matsuda2018} in Mott-insulating Cs$_3$C$_{60}$ using the same vibronic coupling parameters.
Furthermore, vibronic coupling parameters from B3LYP calculations tend to give good description of inelastic electron tunneling spectra of other organic molecule \cite{Shizu2010}.
All these facts show that B3LYP values are closer to the reality in C$_{60}^{+}$, but there still a mismatch when compared with experimental data. 
The application of CAM-B3LYP could eliminate such mismatch of vibronic parameters in fullerene system, as shown in recent study of C$_{60}^-$\cite{Huang_2019_C60t1g}.

%
Although the derivation of vibronic coupling parameters is the first step toward full description of the molecular states of C$_{60}^+$, we believe this is a crucial step.
Once calculations of accurate enough vibronic states become possible, it is possible to interpret various spectra such as scanning tunneling measurements of C$_{60}$ \cite{Frederiksen2008}, inverse PES \cite{Grose2017} and angle resolved PES \cite{PES_C60_2019} to mention a few.
Furthermore, present coupling parameters are derived based on the well-defined phase factor of normal modes which has been also used for orbital coupling parameters of LUMO \cite{Liu2018b} and next LUMO \cite{Huang_2019_C60t1g}.
Therefore, by combining present coupling parameters with them, it is also possible to address complex vibronic problems of excited C$_{60}$ \cite{Qiu_Ceulemans_2001}, and also to analyze \textit{e.g.} luminescence spectra \cite{Akimoto2002} and relaxation process and thermally activated delayed luminescence \cite{Stepanov2002}.

\section{Conclusions}
In this work, orbital vibronic coupling parameters for $h_u$ HOMO level of C$_{60}$ are derived using both B3LYP and CAM-B3LYP hybrid functional.
We believe that these vibronic coupling parameters are high accurate and close to the real situation.
With these obtained coupling parameters, JT stabilization energies of C$_{60}^+$ are calculated, and JT structure at the minima of APES is confirmed to be $D_{5d}$, with the stablization energy 110 meV and 129 meV calculated with B3LYP and CAM-B3LYP, respectively.
JT stabilization energies in C$_{60}^+$ are about two times larger than that in C$_{60}^-$, suggesting the crucial role of the dynamical JT effect to reveal the actual situation of C$_{60}^+$.

Present coupling parameters have been derived within the same framework used for our studies on ground and excited C$_{60}^-$.
Thus, combining present data with that from other works, it is also possible to analyze the vibronic problems of excited C$_{60}$.

\section*{acknowledgments}
The authors thank Dr. Naoya Iwahara and Prof. Dr. Liviu Chibotaru for fruitful discussions.
They also gratefully acknowledge funding by the China Scholarship Council (CSC).
Dr. Dan. Liu is supported by "the Fundamental Research Funds for the Central Universities"(G2019KY0517, G2019KY05104)

\section*{conflict of interest}
There are no conflicts to declare.

\section{appendix}
\subsection{Clebsch-Gordan coefficients: $\hat{C}_{\Gamma\gamma}$}
For the derivation of the vibronic Hamiltonian, the Clebsch-Gordan coefficients, $\hat{C}_{\Gamma\gamma}$ are taken from Ref. \cite{Liu2018b}, and listed below, in which $\phi = (1+\sqrt{5})/2$, $\Gamma$ = g, 1h, 2h, and $\gamma$ = a, x, y, z, $\theta$, $\epsilon$, $\xi$, $\eta$, $\zeta$.

\begin{eqnarray}
\hat{C}_{ga}= \begin{pmatrix}
  \frac{1}{2}\sqrt{\frac{3}{2}}  & 0 & 0 & 0 & 0 \\
  0 & \frac{1}{2}\sqrt{\frac{3}{2}}  & 0 & 0 & 0 \\
  0 & 0 & -\frac{1}{\sqrt{6}} & 0 & 0 \\
  0 & 0 & 0 & -\frac{1}{\sqrt{6}} & 0 \\
  0 & 0 & 0 & 0 & -\frac{1}{\sqrt{6}}
 \end{pmatrix}
\end{eqnarray}

\begin{eqnarray}
\hat{C}_{gx}=  \begin{pmatrix}
  0 & 0 & -\frac{\phi}{4}\sqrt{\frac{5}{2}}  & 0 & 0 \\
  0 & 0 & \frac{\phi^{-2}}{4}\sqrt{\frac{5}{6}} & 0 & 0 \\
 -\frac{\phi}{4}\sqrt{\frac{5}{2}} & \frac{\phi^{-2}}{4}\sqrt{\frac{5}{6}} & 0 & 0 & 0 \\
  0 & 0 & 0 & 0 & \frac{1}{2}\sqrt{\frac{5}{6}} \\
  0 & 0 & 0 & \frac{1}{2}\sqrt{\frac{5}{6}} & 0
 \end{pmatrix}
\end{eqnarray}

\begin{eqnarray}
\hat{C}_{gy}=  \begin{pmatrix}
  0 & 0 & 0 & \frac{\phi^{-1}}{4}\sqrt{\frac{5}{2}} & 0 \\
  0 & 0 & 0 & -\frac{\phi^2}{4}\sqrt{\frac{5}{6}}  & 0 \\
  0 & 0 & 0 & 0 & \frac{1}{2}\sqrt{\frac{5}{6}} \\
 \frac{\phi^{-1}}{4}\sqrt{\frac{5}{2}} & -\frac{\phi^2}{4}\sqrt{\frac{5}{6}} & 0 & 0 & 0 \\
  0 & 0 & \frac{1}{2}\sqrt{\frac{5}{6}} & 0 & 0
 \end{pmatrix}
\end{eqnarray}

\begin{eqnarray}
\hat{C}_{gz}=  \begin{pmatrix}
  0 & 0 & 0 & 0 & \frac{1}{4}\sqrt{\frac{5}{2}} \\
  0 & 0 & 0 & 0 & \frac{5}{4\sqrt{6}}  \\
  0 & 0 & 0 & \frac{1}{2}\sqrt{\frac{5}{6}} & 0 \\
  0 & 0 & \frac{1}{2}\sqrt{\frac{5}{6}} & 0 & 0 \\
 \frac{1}{4}\sqrt{\frac{5}{2}} & \frac{5}{4\sqrt{6}} & 0 & 0 & 0
 \end{pmatrix}
\end{eqnarray}

\begin{eqnarray}
\hat{C}_{1h\theta} = \begin{pmatrix}
  \frac{\sqrt{5}}{16} & \frac{3\sqrt{3}}{16} & 0 & 0 & 0 \\
  \frac{3\sqrt{3}}{16} & -\frac{\sqrt{5}}{16} & 0 & 0 & 0 \\
  0 & 0 & -\frac{\phi^{-2}}{4} & 0 & 0 \\
  0 & 0 & 0 & \frac{\phi^2}{4} & 0 \\
  0 & 0 & 0 & 0 & -\frac{\sqrt{5}}{4}
 \end{pmatrix}
\end{eqnarray}

\begin{eqnarray}
\hat{C}_{1h\epsilon} = \begin{pmatrix}
  \frac{3\sqrt{3}}{16} & -\frac{\sqrt{5}}{16} & 0 & 0 & 0 \\
  -\frac{\sqrt{5}}{16} & -\frac{3\sqrt{3}}{16} & 0 & 0 & 0 \\
  0 & 0 & -\frac{\sqrt{3}\phi}{4} & 0 & 0 \\
  0 & 0 & 0 & \frac{\sqrt{3}\phi^{-1}}{4} & 0 \\
  0 & 0 & 0 & 0 & \frac{\sqrt{3}}{4}
 \end{pmatrix}
\end{eqnarray}

\begin{eqnarray}
\hat{C}_{1h\xi} = \begin{pmatrix}
  0 & 0 & -\frac{\phi^{-2}}{4} & 0 & 0 \\
  0 & 0 & -\frac{\sqrt{3}\phi}{4} & 0 & 0 \\
  -\frac{\phi^{-2}}{4} & -\frac{\sqrt{3}\phi}{4} & 0 & 0 & 0 \\
  0 & 0 & 0 & 0 & 0 \\
  0 & 0 & 0 & 0 & 0
 \end{pmatrix}
\end{eqnarray}

\begin{eqnarray}
\hat{C}_{1h\eta} = \begin{pmatrix}
  0 & 0 & 0 & \frac{\phi^2}{4} & 0 \\
  0 & 0 & 0 & \frac{\sqrt{3}\phi^{-1}}{4}& 0 \\
  0 & 0 & 0 & 0 & 0 \\
  \frac{\phi^2}{4} & \frac{\sqrt{3}\phi^{-1}}{4} & 0 & 0 & 0 \\
  0 & 0 & 0 & 0 & 0
 \end{pmatrix}
\end{eqnarray}

\begin{eqnarray}
\hat{C}_{1h\zeta} = \begin{pmatrix}
  0 & 0 & 0 & 0 & -\frac{\sqrt{5}}{4} \\
  0 & 0 & 0 & 0 & \frac{\sqrt{3}}{4} \\
  0 & 0 & 0 & 0 & 0 \\
  0 & 0 & 0 & 0 & 0 \\
 -\frac{\sqrt{5}}{4} & \frac{\sqrt{3}}{4}& 0 & 0 & 0
 \end{pmatrix}
\end{eqnarray}

\begin{eqnarray}
\hat{C}_{2h\theta} = \begin{pmatrix}
  \frac{9}{16} & -\frac{\sqrt{15}}{16} & 0 & 0 & 0 \\
  -\frac{\sqrt{15}}{16} & -\frac{9}{16} & 0 & 0 & 0 \\
  0 & 0 & \frac{\phi}{4} & 0 & 0 \\
  0 & 0 & 0 & -\frac{\phi^{-1}}{4} & 0 \\
  0 & 0 & 0 & 0 & -\frac{1}{4}
 \end{pmatrix}
\end{eqnarray}

\begin{eqnarray}
\hat{C}_{2h\epsilon} = \begin{pmatrix}
  -\frac{\sqrt{15}}{16} & -\frac{9}{16} & 0 & 0 & 0  \\
  -\frac{9}{16} & \frac{\sqrt{15}}{16} & 0 & 0 & 0  \\
  0 & 0 & -\frac{\phi^{-2}}{4\sqrt{3}} & 0 & 0  \\
  0 & 0 & 0 & \frac{\phi^2}{4\sqrt{3}} & 0 \\
  0 & 0 & 0 & 0 & -\frac{1}{4}\sqrt{\frac{5}{3}}
 \end{pmatrix}
\end{eqnarray}

\begin{eqnarray}
\hat{C}_{2h\xi} = \begin{pmatrix}
  0 & 0 & \frac{\phi}{4} & 0 & 0  \\
  0 & 0 & -\frac{\phi^{-2}}{4\sqrt{3}} & 0 & 0  \\
  \frac{\phi}{4} & -\frac{\phi^{-2}}{4\sqrt{3}} & 0 & 0 & 0 \\
  0 & 0 & 0 & 0 & \frac{1}{\sqrt{3}} \\
  0 & 0 & 0 & \frac{1}{\sqrt{3}} & 0
 \end{pmatrix}
\end{eqnarray}

\begin{eqnarray}
\hat{C}_{2h\eta} = \begin{pmatrix}
  0 & 0 & 0 & -\frac{\phi^{-1}}{4} & 0 \\
  0 & 0 & 0 & \frac{\phi^2}{4\sqrt{3}} & 0 \\
  0 & 0 & 0 & 0 & \frac{1}{\sqrt{3}} \\
 -\frac{\phi^{-1}}{4} & \frac{\phi^2}{4\sqrt{3}} & 0 & 0 & 0  \\
  0 & 0 & \frac{1}{\sqrt{3}} & 0 & 0
 \end{pmatrix}
\end{eqnarray}

\begin{eqnarray}
\hat{C}_{2h\zeta} = \begin{pmatrix}
  0 & 0 & 0 & 0 & -\frac{1}{4} \\
  0 & 0 & 0 & 0 & -\frac{1}{4}\sqrt{\frac{5}{3}} \\
  0 & 0 & 0 & \frac{1}{\sqrt{3}} & 0 \\
  0 & 0& \frac{1}{\sqrt{3}} & 0 & 0  \\
  -\frac{1}{4} & -\frac{1}{4}\sqrt{\frac{5}{3}} & 0 & 0 & 0
 \end{pmatrix}
\end{eqnarray}

\bibliography{sample}

\clearpage
\newpage
\onecolumngrid
\begin{center}
  \textbf{Supplement Material: First principles study of the vibronic coupling in positively charged C$_{60}^+$}\\[.2cm]
  Zhishuo Huang$^{1,a)}$ and Dan Liu$^{2,1,b)}$\\[.1cm]
  {\itshape ${}^1$Theory of Nanomaterials Group, KU Leuven, Celestijnenlaan 200F, B-3001 Leuven, Belgium\\
  ${}^2$Institute of Flexible Electronics (IFE), Northwestern Polytechnical University, 127 West Youyi Road, Xi'an, 710072, Shaanxi, China\\}
  ${}^{a)}$Electronic address: zhishuohuang@gmail.com\\
  ${}^{b)}$Electronic address: iamdliu@nwpu.edu.cn
\end{center}
\date{\today}

  Supplemental Materials contain the JT splitting of the HOMO levels with respect to $q_{h\epsilon}$ and $q_{ga}$deformations.

\section{JT splitting of the HOMO levels}

There are eight $q_{h\epsilon}$  and six $q_{ga}$ deformations, which are distinguished by the subindex, as $q_{h(1)\epsilon}$ corresponding to the first $q_{h\epsilon}$ deformation. 

\begin{figure}[bt]
\includegraphics[width=10cm]{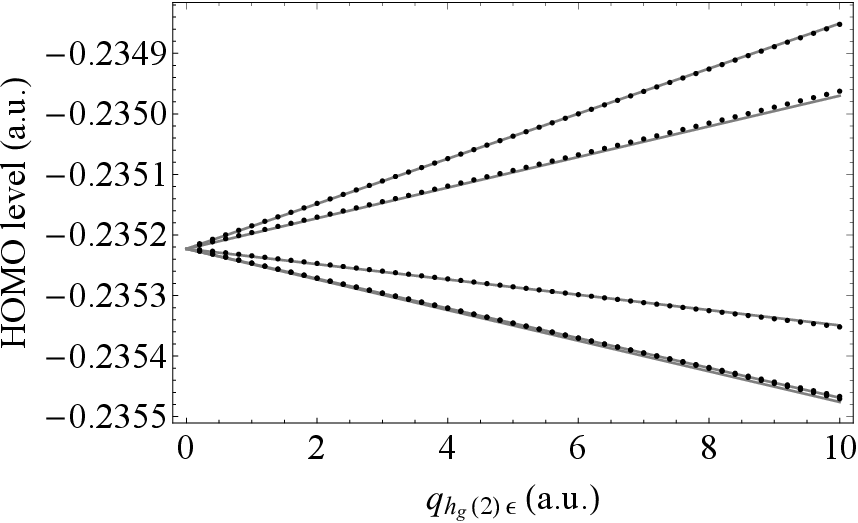}
\caption{
The JT splitting of the HOMO levels calculated by B3LYP with respect to $q_{h_g(2)\epsilon}$ deformation (in atomic unit).
The black points and gray lines indicate the DFT values and model energy, respectively.
}
\label{Fig:Vh2}
\end{figure}

\begin{figure}[bt]
\includegraphics[width=10cm]{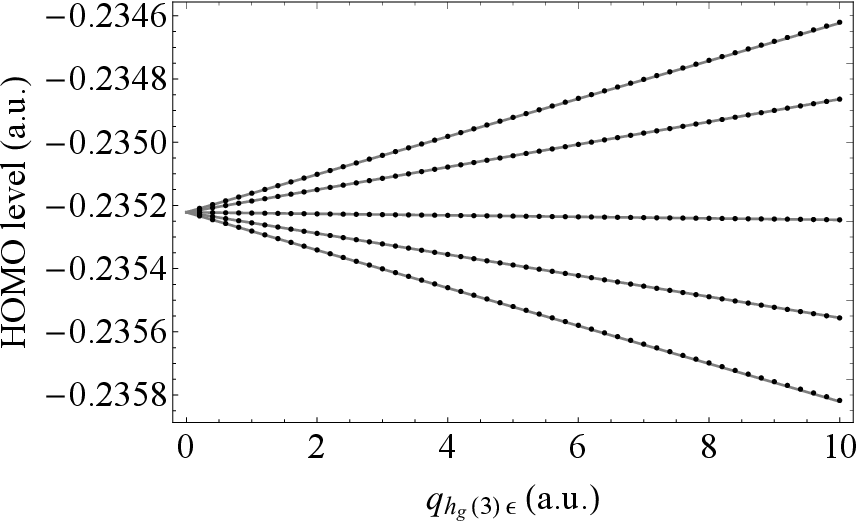}
\caption{
The JT splitting of the HOMO levels calculated by B3LYP with respect to $q_{h_g(3)\epsilon}$ deformation (in atomic unit).
The black points and gray lines indicate the DFT values and model energy, respectively.
}
\label{Fig:Vh3}
\end{figure}

\begin{figure}[bt]
\includegraphics[width=10cm]{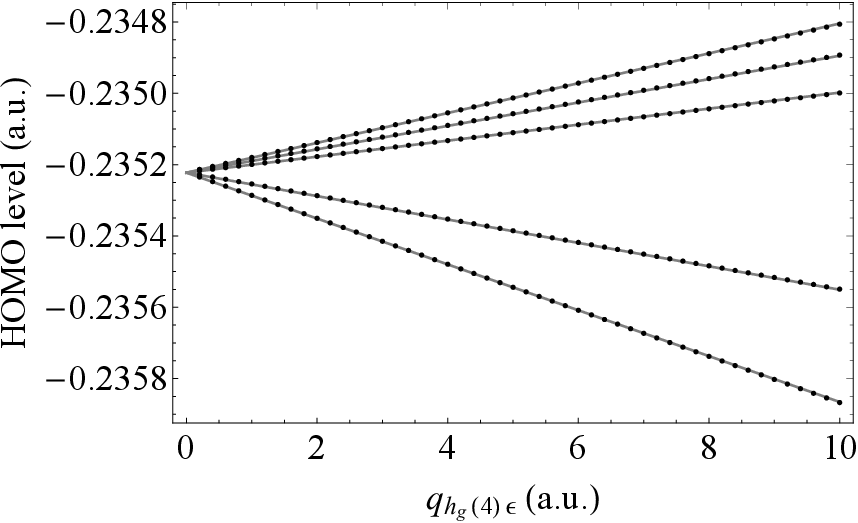}
\caption{
The JT splitting of the HOMO levels calculated by B3LYP with respect to $q_{h_g(4)\epsilon}$ deformation (in atomic unit).
The black points and gray lines indicate the DFT values and model energy, respectively.
}
\label{Fig:Vh4}
\end{figure}

\begin{figure}[bt]
\includegraphics[width=10cm]{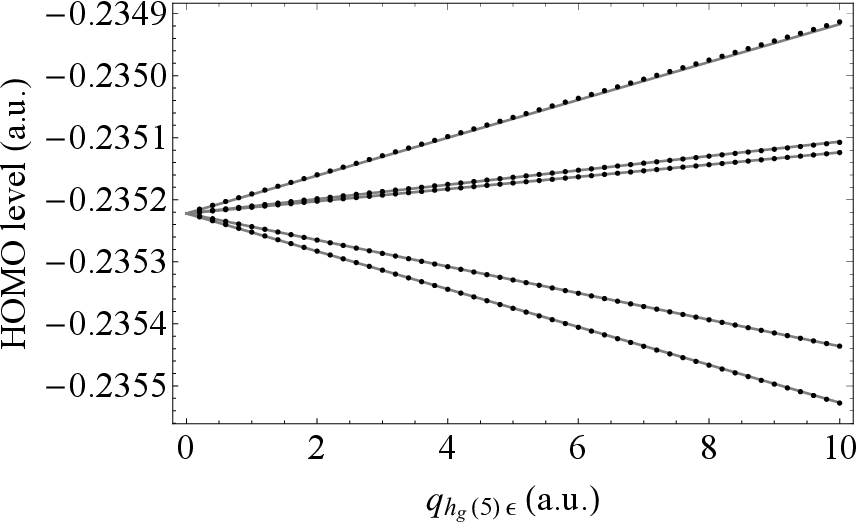}
\caption{
The JT splitting of the HOMO levels calculated by B3LYP with respect to $q_{h_g(5)\epsilon}$ deformation (in atomic unit).
The black points and gray lines indicate the DFT values and model energy, respectively.
}
\label{Fig:Vh5}
\end{figure}

\begin{figure}[bt]
\includegraphics[width=10cm]{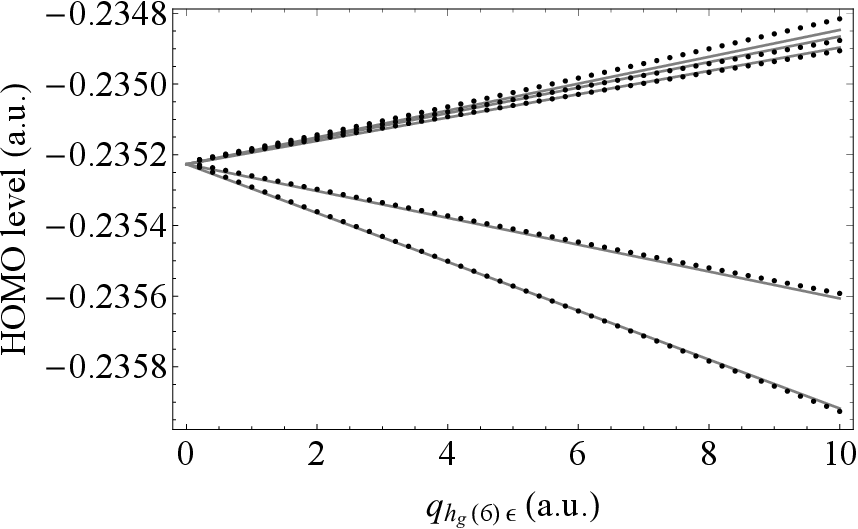}
\caption{
The JT splitting of the HOMO levels calculated by B3LYP with respect to $q_{h_g(6)\epsilon}$ deformation (in atomic unit).
The black points and gray lines indicate the DFT values and model energy, respectively.
}
\label{Fig:Vh6}
\end{figure}

\begin{figure}[bt]
\includegraphics[width=10cm]{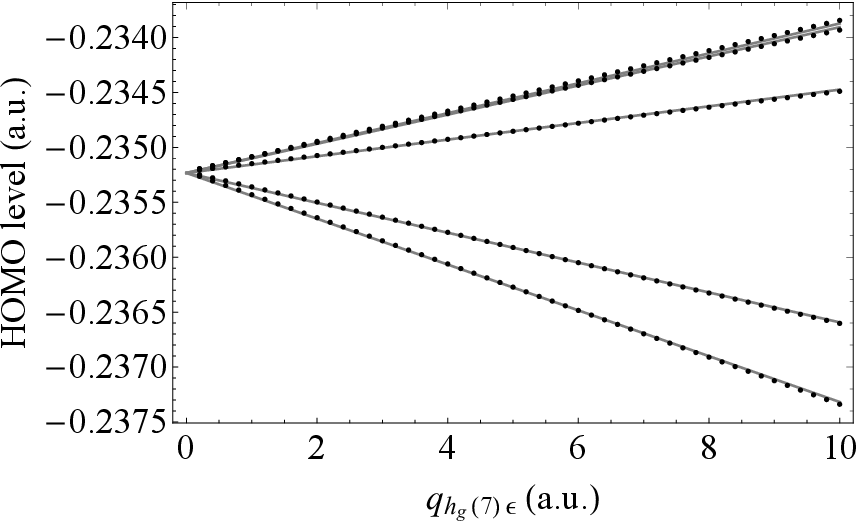}
\caption{
The JT splitting of the HOMO levels calculated by B3LYP with respect to $q_{h_g(7)\epsilon}$ deformation (in atomic unit).
The black points and gray lines indicate the DFT values and model energy, respectively.
}
\label{Fig:Vh7}
\end{figure}

\begin{figure}[bt]
\includegraphics[width=10cm]{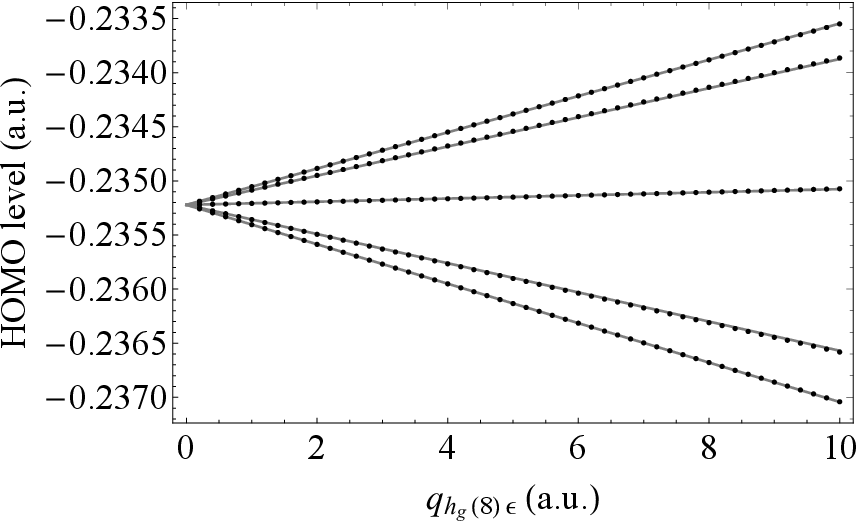}
\caption{
The JT splitting of the HOMO levels calculated by B3LYP with respect to $q_{h_g(8)\epsilon}$ deformation (in atomic unit).
The black points and gray lines indicate the DFT values and model energy, respectively.
}
\label{Fig:Vh8}
\end{figure}

\begin{figure}[bt]
\includegraphics[width=10cm]{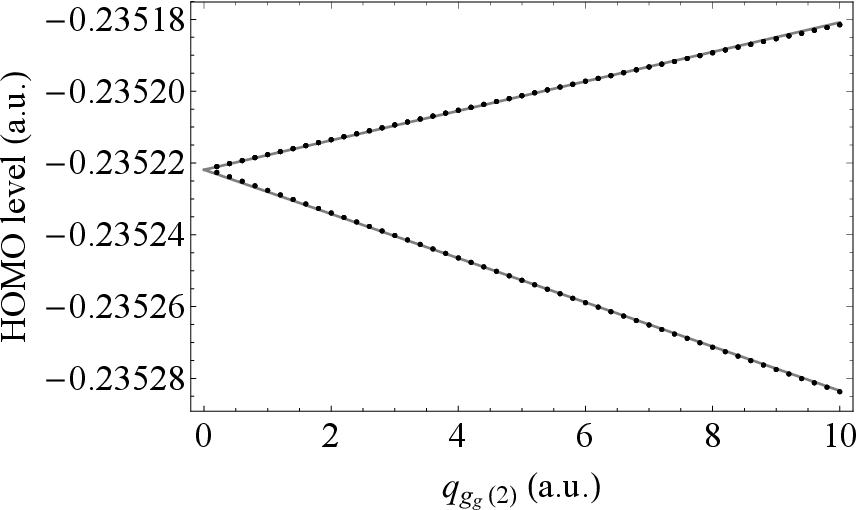}
\caption{
The JT splitting of the HOMO levels calculated by B3LYP with respect to $q_{g_g(2)\epsilon}$ deformation (in atomic unit).
The black points and gray lines indicate the DFT values and model energy, respectively.
}
\label{Fig:Vg2}
\end{figure}

\begin{figure}[bt]
\includegraphics[width=10cm]{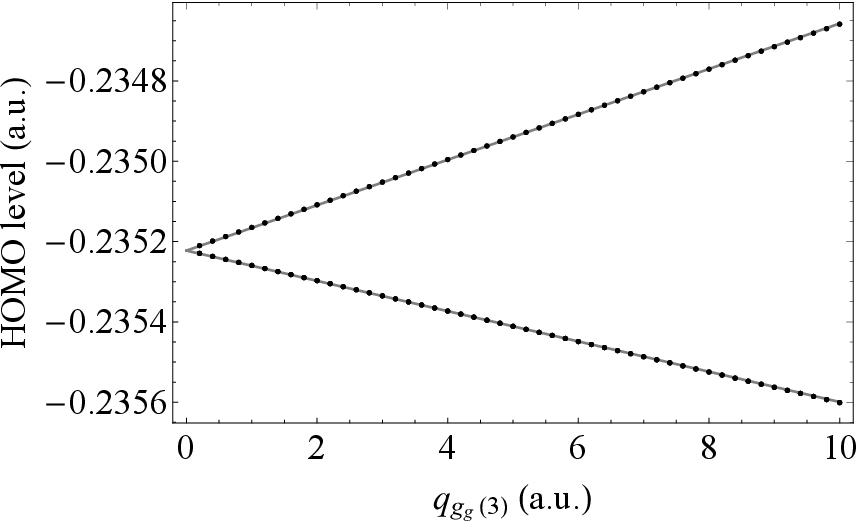}
\caption{
The JT splitting of the HOMO levels calculated by B3LYP with respect to $q_{g_g(3)\epsilon}$ deformation (in atomic unit).
The black points and gray lines indicate the DFT values and model energy, respectively.
}
\label{Fig:Vg3}
\end{figure}

\begin{figure}[bt]
\includegraphics[width=10cm]{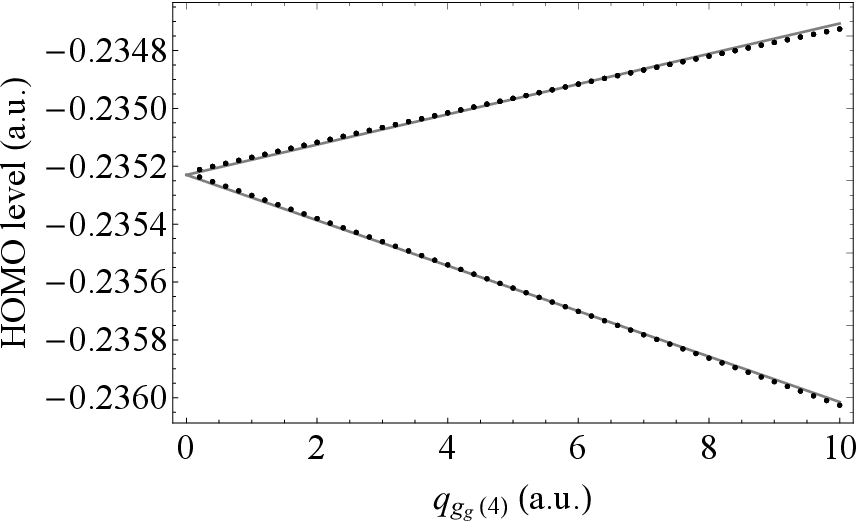}
\caption{
The JT splitting of the HOMO levels calculated by B3LYP with respect to $q_{g_g(4)\epsilon}$ deformation (in atomic unit).
The black points and gray lines indicate the DFT values and model energy, respectively.
}
\label{Fig:Vg4}
\end{figure}

\begin{figure}[bt]
\includegraphics[width=10cm]{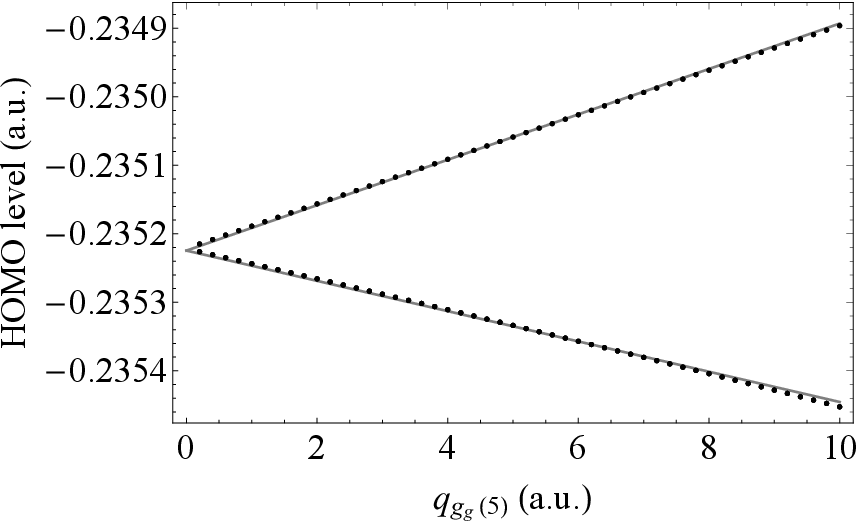}
\caption{
The JT splitting of the HOMO levels calculated by B3LYP with respect to $q_{g_g(5)\epsilon}$ deformation (in atomic unit).
The black points and gray lines indicate the DFT values and model energy, respectively.
}
\label{Fig:Vg5}
\end{figure}

\begin{figure}[bt]
\includegraphics[width=10cm]{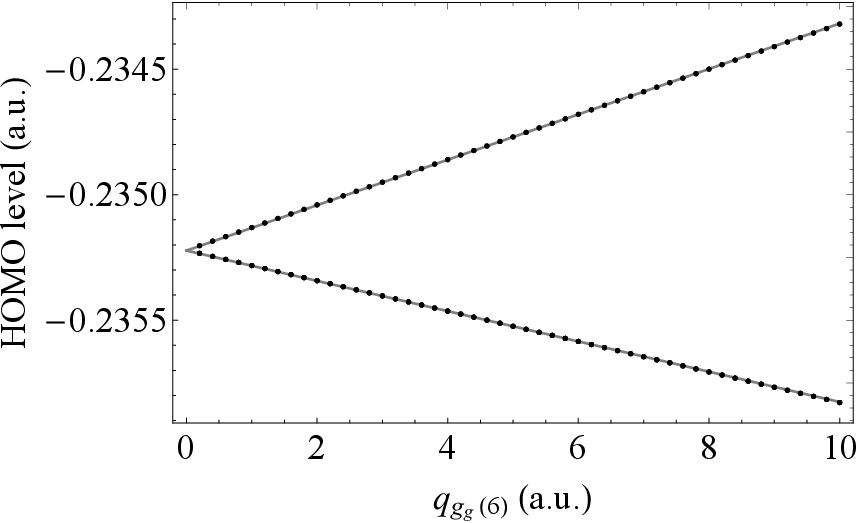}
\caption{
The JT splitting of the HOMO levels calculated by B3LYP with respect to $q_{g_g(6)\epsilon}$ deformation (in atomic unit).
The black points and gray lines indicate the DFT values and model energy, respectively.
}
\label{Fig:Vg6}
\end{figure}

The DFT data with B3LYP hybrid functional and the defination of the phase factors of the normal modes are taken from Ref. \cite{Liu2018b}.
The fitting for of the DFT HOMO levels to the model hamiltonian for $q_{h_g(i)\epsilon},~i=2,3,4,5,6,7,8$ are shown in Fig. \ref{Fig:Vh2}, \ref{Fig:Vh3}, \ref{Fig:Vh4}, \ref{Fig:Vh5}, \ref{Fig:Vh6}, \ref{Fig:Vh7} and \ref{Fig:Vh8}, 
while $q_{g_g(i)},~i=2,3,4,5,6$ are shown in Fig. \ref{Fig:Vg2}, \ref{Fig:Vg3}, \ref{Fig:Vg4}, \ref{Fig:Vg5}, and \ref{Fig:Vg6}.

\clearpage

\clearpage

\begin{figure}[bt]
\includegraphics[width=10cm]{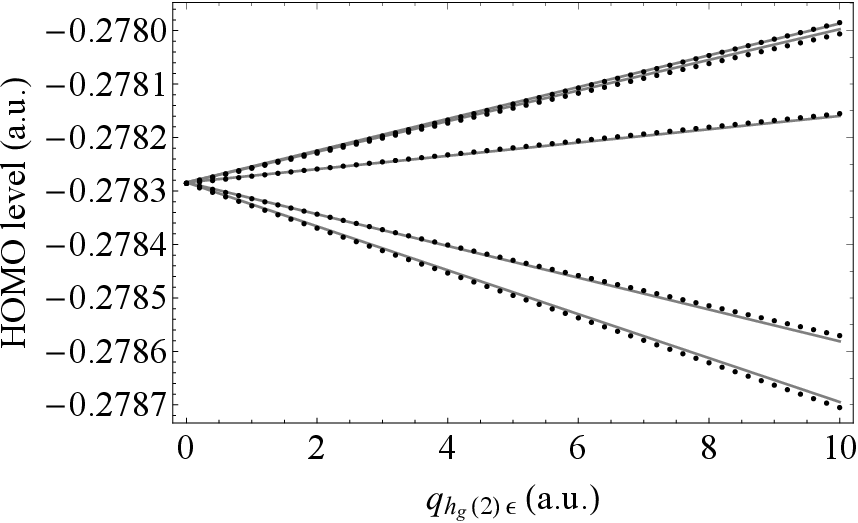}
\caption{
The JT splitting of the HOMO levels calculated by CAM-B3LYP with respect to $q_{h_g(2)\epsilon}$ deformation (in atomic unit).
The black points and gray lines indicate the DFT values and model energy, respectively.
}
\label{Fig:CAM_Vh2}
\end{figure}

\begin{figure}[bt]
\includegraphics[width=10cm]{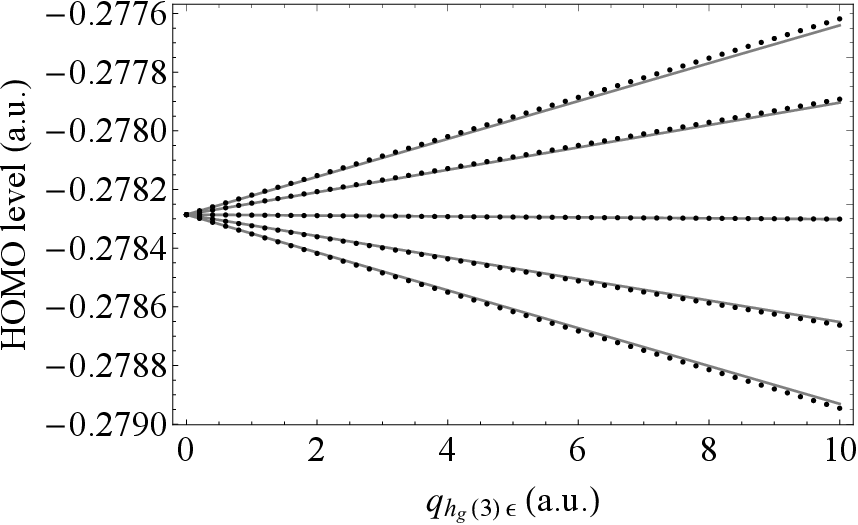}
\caption{
The JT splitting of the HOMO levels calculated by CAM-B3LYP with respect to $q_{h_g(3)\epsilon}$ deformation (in atomic unit).
The black points and gray lines indicate the DFT values and model energy, respectively.
}
\label{Fig:CAM_Vh3}
\end{figure}

\begin{figure}[bt]
\includegraphics[width=10cm]{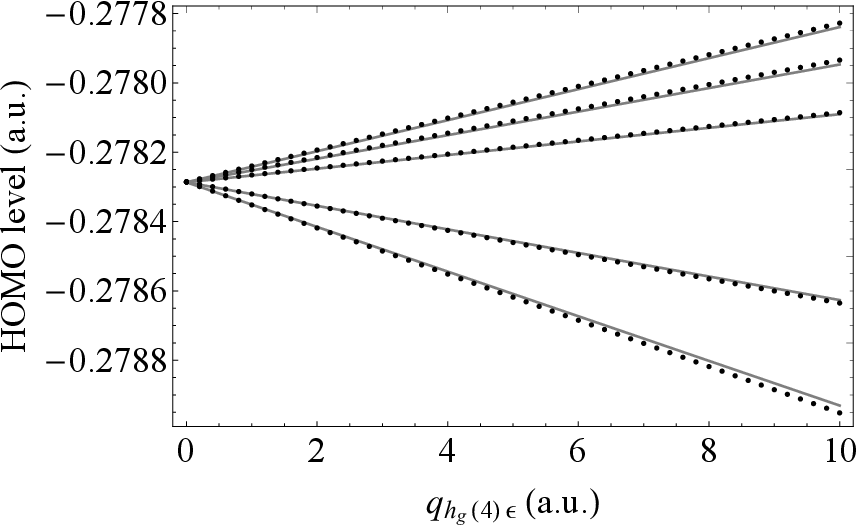}
\caption{
The JT splitting of the HOMO levels calculated by CAM-B3LYP with respect to $q_{h_g(4)\epsilon}$ deformation (in atomic unit).
The black points and gray lines indicate the DFT values and model energy, respectively.
}
\label{Fig:CAM_Vh4}
\end{figure}

\begin{figure}[bt]
\includegraphics[width=10cm]{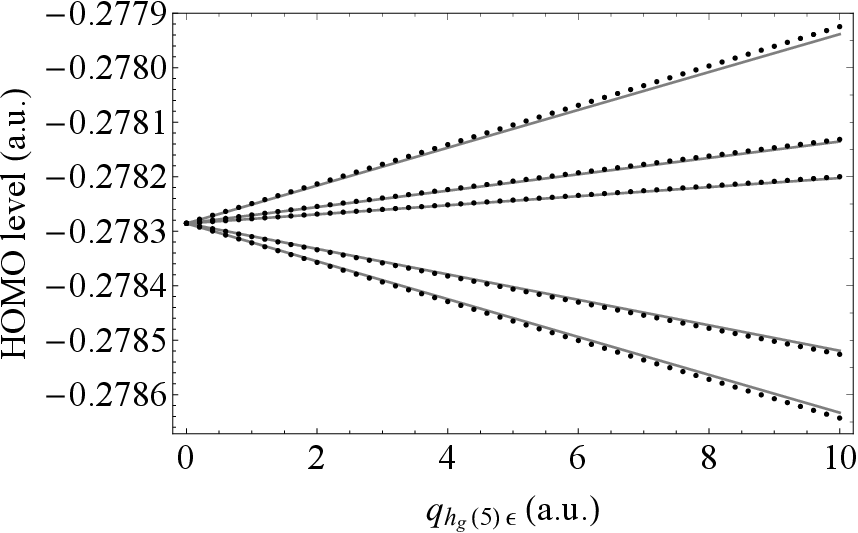}
\caption{
The JT splitting of the HOMO levels calculated by CAM-B3LYP with respect to $q_{h_g(5)\epsilon}$ deformation (in atomic unit).
The black points and gray lines indicate the DFT values and model energy, respectively.
}
\label{Fig:CAM_Vh5}
\end{figure}

\begin{figure}[bt]
\includegraphics[width=10cm]{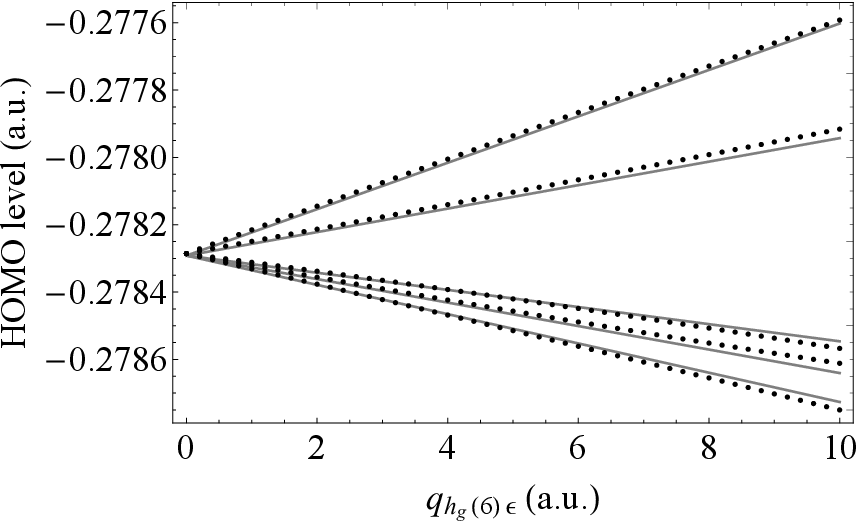}
\caption{
The JT splitting of the HOMO levels calculated by CAM-B3LYP with respect to $q_{h_g(6)\epsilon}$ deformation (in atomic unit).
The black points and gray lines indicate the DFT values and model energy, respectively.
}
\label{Fig:CAM_Vh6}
\end{figure}

\begin{figure}[bt]
\includegraphics[width=10cm]{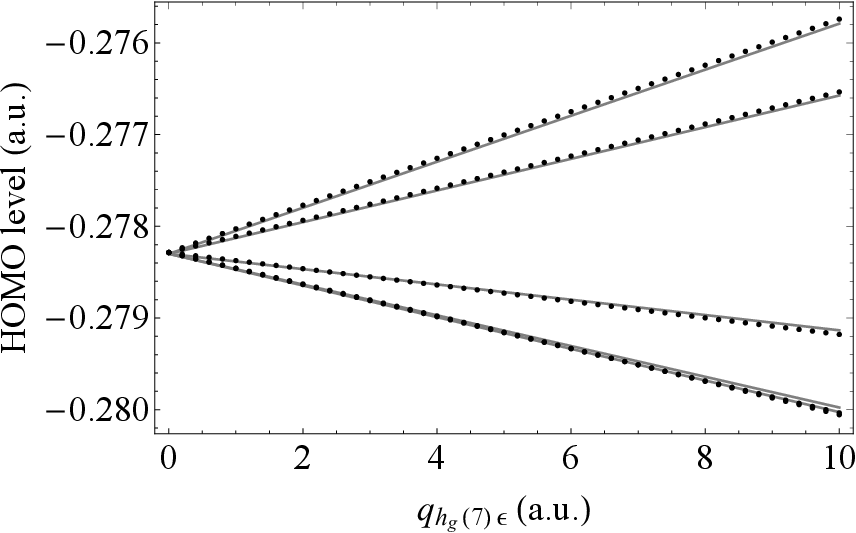}
\caption{
The JT splitting of the HOMO levels calculated by CAM-B3LYP with respect to $q_{h_g(7)\epsilon}$ deformation (in atomic unit).
The black points and gray lines indicate the DFT values and model energy, respectively.
}
\label{Fig:CAM_Vh7}
\end{figure}

\begin{figure}[bt]
\includegraphics[width=10cm]{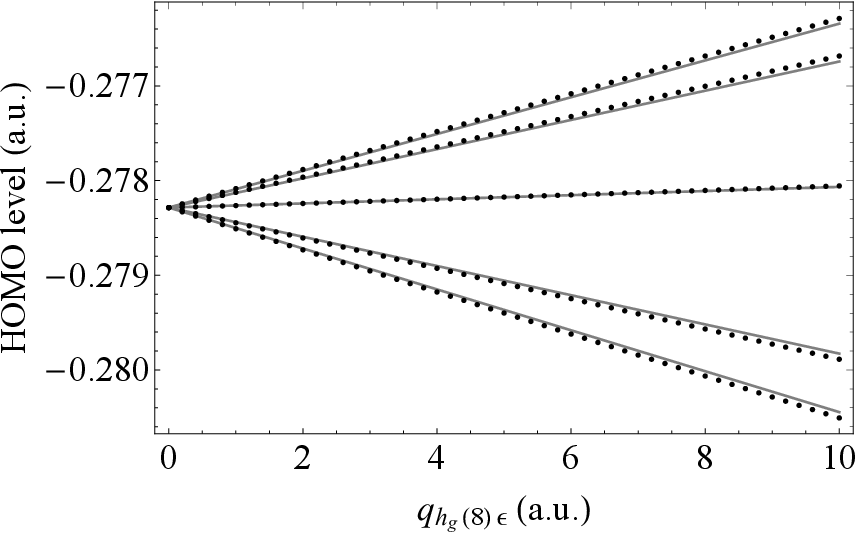}
\caption{
The JT splitting of the HOMO levels calculated by CAM-B3LYP with respect to $q_{h_g(8)\epsilon}$ deformation (in atomic unit).
The black points and gray lines indicate the DFT values and model energy, respectively.
}
\label{Fig:CAM_Vh8}
\end{figure}

\begin{figure}[bt]
\includegraphics[width=10cm]{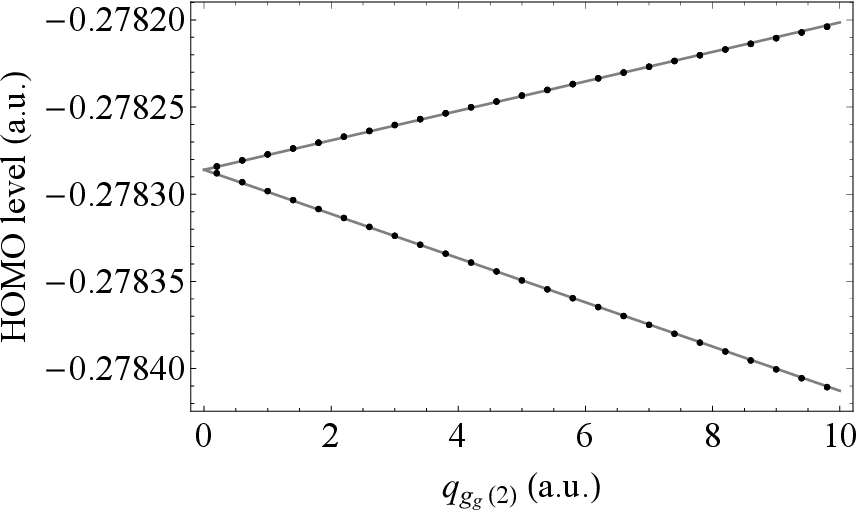}
\caption{
The JT splitting of the HOMO levels calculated by CAM-B3LYP with respect to $q_{g_g(2)\epsilon}$ deformation (in atomic unit).
The black points and gray lines indicate the DFT values and model energy, respectively.
}
\label{Fig:CAM_Vg2}
\end{figure}

\begin{figure}[bt]
\includegraphics[width=10cm]{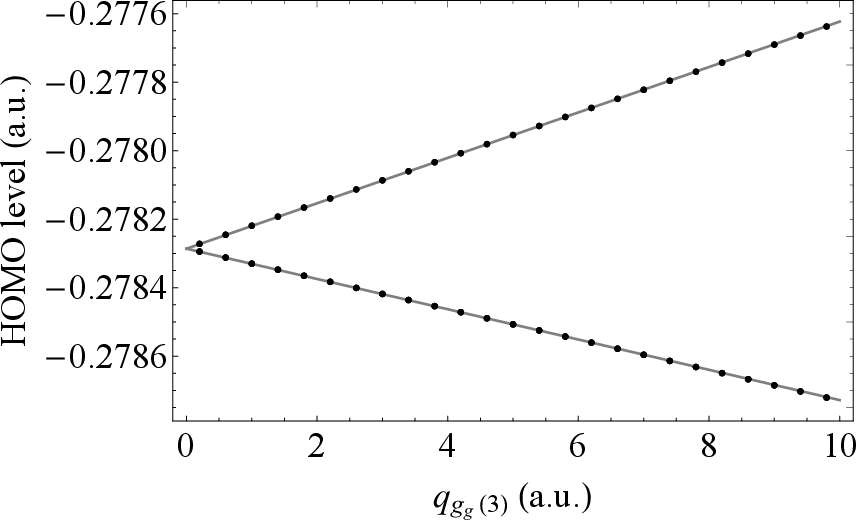}
\caption{
The JT splitting of the HOMO levels calculated by CAM-B3LYP with respect to $q_{g_g(3)\epsilon}$ deformation (in atomic unit).
The black points and gray lines indicate the DFT values and model energy, respectively.
}
\label{Fig:CAM_Vg3}
\end{figure}

\begin{figure}[bt]
\includegraphics[width=10cm]{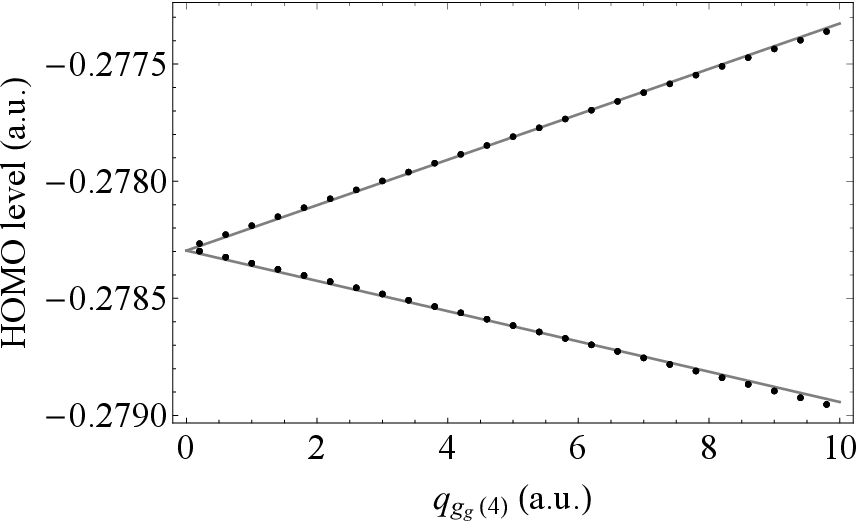}
\caption{
The JT splitting of the HOMO levels calculated by CAM-B3LYP with respect to $q_{g_g(4)\epsilon}$ deformation (in atomic unit).
The black points and gray lines indicate the DFT values and model energy, respectively.
}
\label{Fig:CAM_Vg4}
\end{figure}

\begin{figure}[bt]
\includegraphics[width=10cm]{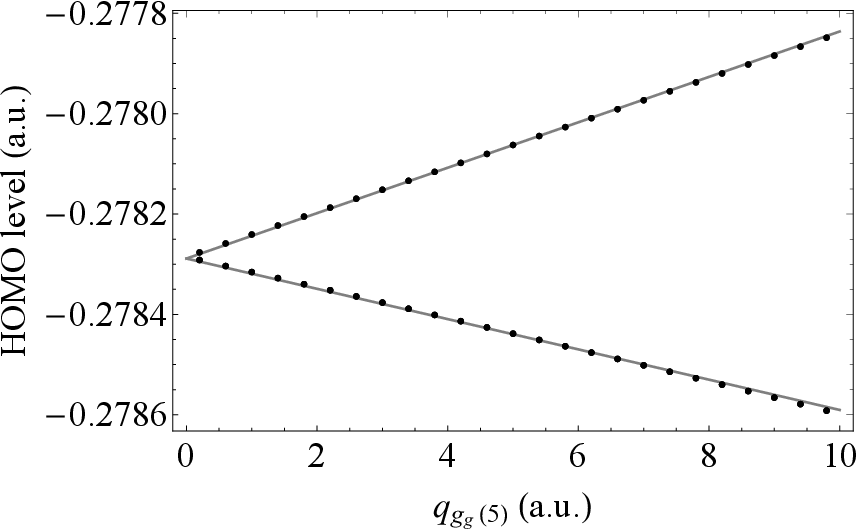}
\caption{
The JT splitting of the HOMO levels calculated by CAM-B3LYP with respect to $q_{g_g(5)\epsilon}$ deformation (in atomic unit).
The black points and gray lines indicate the DFT values and model energy, respectively.
}
\label{Fig:CAM_Vg5}
\end{figure}

\begin{figure}[bt]
\includegraphics[width=10cm]{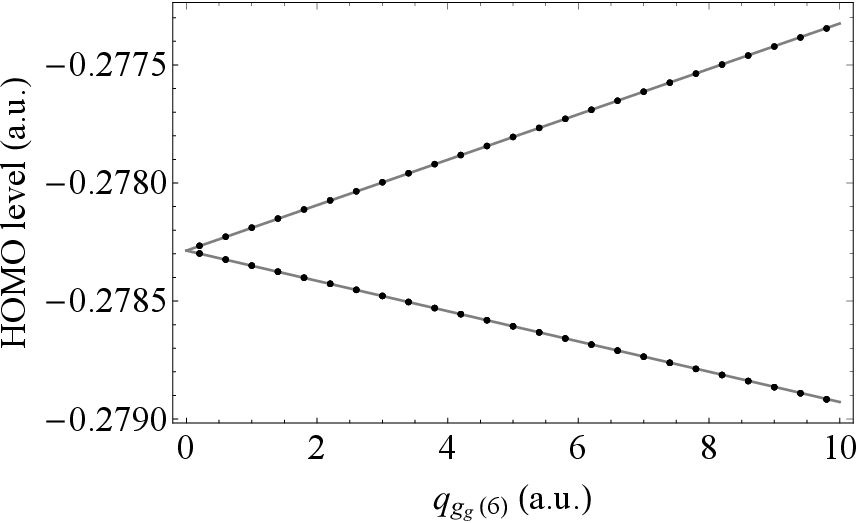}
\caption{
The JT splitting of the HOMO levels calculated by CAM-B3LYP with respect to $q_{g_g(6)\epsilon}$ deformation (in atomic unit).
The black points and gray lines indicate the DFT values and model energy, respectively.
}
\label{Fig:CAM_Vg6}
\end{figure}

Considering CAM-B3LYP functional, Fig. \ref{Fig:CAM_Vh2}, \ref{Fig:CAM_Vh3}, \ref{Fig:CAM_Vh4}, \ref{Fig:CAM_Vh5}, \ref{Fig:CAM_Vh6}, \ref{Fig:CAM_Vh7} and \ref{Fig:CAM_Vh8} depict the fitting of $q_{h_g(i)\epsilon},~i=2,3,4,5,6,7,8$,
with Fig. \ref{Fig:CAM_Vg2}, \ref{Fig:CAM_Vg3}, \ref{Fig:CAM_Vg4}, \ref{Fig:CAM_Vg5}, and \ref{Fig:CAM_Vg6} for $q_{g_g(i)},~i=2,3,4,5,6$.

\end{document}